\documentclass[aps,prper,reprint, longbibliography, showkeys, nofootinbib]{revtex4-2}

\usepackage[T1]{fontenc}	
\usepackage{geometry}		
\usepackage{graphicx}
\usepackage{epsfig}
\usepackage[above,below]{placeins}	
\usepackage{times}
\usepackage{csquotes} 
\usepackage{gensymb} 

\usepackage{enumitem}          
\setlist{nosep} 

\usepackage{hyperref}  
\hypersetup{colorlinks=true,urlcolor=blue,citecolor=blue,linkcolor=blue}  

\begin{document}
\title{Dynamics of Productive Confirmation Framing in an Introductory Lab}

\author{Ian Descamps$^{1}$, Sophia Jeon$^{2}$, N. G. Holmes$^{3}$, Rachel E. Scherr$^{4}$, and David Hammer$^{1,2}$}
\email[Please address correspondence to ]{ian.descamps@tufts.com}
\affiliation{$^1$Department of Education, Tufts University, 12 Upper Campus Road, Medford, Massachusetts 02155, USA\\
$^2$Department of Physics \& Astronomy, Tufts University,
574 Boston Avenue, Suite 304, Medford, Massachusetts 02155, USA\\ $^3$Laboratory of Atomic and Solid State Physics, Cornell University, Ithaca, New York 14853, USA \\ $^4$Physical Science Division, School of STEM, University of Washington Bothell,
Bothell, Washington 98011, USA}

\begin{abstract}
\label{abstract}

In introductory physics laboratory instruction, students often expect to confirm or demonstrate textbook physics concepts \cite{wilcox_students_2017, Hu_Zwickl_2017, hu_examining_2018-1}. 
This expectation is largely undesirable: labs that emphasize confirmation of textbook physics concepts are unsuccessful at teaching those concepts \cite{wieman_measuring_2015, holmes_value_2017} and even in contexts that don't emphasize confirmation, such expectations can lead to students disregarding or manipulating their data in order to obtain the expected result \cite{smith_how_2020}. 
In other words, when students expect their lab activities to confirm a known result, they may relinquish epistemic agency and violate disciplinary practices. 
We claim that, in other cases, confirmatory expectations can actually support productive disciplinary engagement. 
In particular, when an expected result is not confirmed, students may enter a productive "troubleshooting" mode \cite{smith_how_2020}.
We analyze the complex dynamics of students' epistemological framing in a lab where student's confirmatory expectations support and even generate epistemic agency and disciplinary practices, including developing original ideas, measures, and apparatuses to apply to the material world.

\end{abstract}

\maketitle

\section{Introduction}
\label{intro}
In contemporary PER, there are several efforts at reform in introductory physics labs to position students with greater epistemic agency, in line with calls for reform in physics instruction \cite{otero_past_2017, miller_addressing_2018} to "to position [students] as doers of science, rather than receivers of facts" (\cite{miller_addressing_2018} p. 1056). 
In these efforts, the instructional focus shifts from objectives of reinforcing traditional course content toward those of students' productive engagement in experimental activities (e.g, \cite{holmes_value_2017, smith_best_2021}) such as handling measurement uncertainties \cite{kung_teaching_2005}. 
Instructional methods shift as well, from guiding students through reliable procedures toward designing activities that provoke puzzlement and provide opportunities for students to problematize and investigate for themselves. 

One approach to these labs is to design activities that set students up to encounter particular empirical discrepancies. Heckenberg \cite{Heckenber_pendulum}, for example, describes a lab with the "stated aim" of finding the acceleration g to within 1 in $10^4$. Students consider a variety of approaches, generally choosing careful measurement of the period of pendulums. They solve for g from the simple harmonic period $ T = 2\pi \sqrt{\frac{L}{g}}$.
There is direct instruction on the propagation of errors, from the measured quantities L and T (pendulum length and period time, respectively), and on accepted "methodology for the treatment of uncertainties in measurement." 
The true pedagogical aim, however, is that students encounter and contend with the discrepancies that arise, among their measurements and with the expected value, from the divergence of a simple pendulum from simple harmonic oscillation.

Another approach has students measure a genuinely unknown quantity, such the minimum size of a target for a projectile to hit reliably from a launching mechanism they design \cite{kung_teaching_2005}, by methods they devise. 
Project-based approaches have students find their own problems: Phillips \textit{et al.}  \cite{phillips_physicality_2023} discuss a beyond-first-year computational modeling course developed by Tim Atherton at Tufts, with open assignments such as to "make an oscillator" using physical materials, study it, and build a computational model of its motion. 

In this article, we take an approach most comparable to Heckenberg \cite{Heckenber_pendulum}, focusing on labs designed to challenge students with an empirical discrepancy (see also \cite{phillips_not_2021, sundstrom_instructing_2023}. 
Prior work has suggested that students' \textit{epistemological framing} – their expectations about knowledge and learning in labs – plays a key role in whether and how students engage with empirical discrepancies \cite{smith_how_2020, sundstrom_problematizing_2020, phillips_not_2021, sundstrom_instructing_2023}.
Most students in first-year courses expect the purpose of physics experiments to be confirming known results \cite{wilcox_developing_2017, hu_qualitative_2017, hu_examining_2018-1}, even in "reformed" labs that do not prioritize confirmation of textbook concepts \cite{smith_how_2020}.
These expectations are classified as non-expertlike and largely considered undesirable. 

In such a reformed context, Phillips \textit{et al.} \cite{phillips_not_2021} found that many students who framed lab as about confirming a known result failed to notice and attend to empirical discrepancies. 
Moreover, Holmes and her colleagues suggest that such confirmatory expectations are problematic for learning in lab, connecting them with questionable research practices \cite{holmes_doing_2013, bogdan_effects_2013, smith_surprise_2018, stein_confirming_2018, smith_how_2020, sundstrom_instructing_2023}. 
Smith \textit{et al.}  \cite{smith_how_2020} observed that, while students who framed lab as about confirming a known result did notice and attend to empirical discrepancies, they often responded by disregarding or manipulating their data. 
They found that students prioritize "getting correct answers" over staying true to their evidence. 

Smith \textit{et al.} \cite{smith_how_2020} observed students downplay or outright reject the notion that their experimental activity is meaningful and that they can produce knowledge. 
In other words, when students expect their lab activities to confirm a known result, they may relinquish epistemic agency. 
A natural implication of these findings is that instructors (and curriculum designers) should try to shift how students orient towards lab activities to prevent and avoid such confirmation expectations \cite{sundstrom_instructing_2023}. 

Smith \textit{et al.} \cite{smith_how_2020} were careful, however, to note the possibility of confirmatory expectations supporting productive disciplinary engagement:
\begin{displayquote}
"Our data do not allow us to claim that confirmatory expectations necessarily lead students to engage in questionable research practices. For example, not confirming an expected result may suggest an error was made, and productively send the student into a troubleshooting mode" (p. 13)
\end{displayquote}

In this paper, we present an episode of students' "troubleshooting" – and more generally, productive behavior – supported, we claim, by their confirmatory expectations. 
They are epistemically agentive: they take their data seriously as a meaningful reflection of the phenomenon they have constructed. 
Their confirmation framing not only affords but also helps generate productive disciplinary practices as the students enact various (epistemic) actions targeted at reconciling their results and their expectations, including troubleshooting and developing and interpreting graphical representations.

This episode complicates the notion that confirmatory expectations are problematic and naturally raises questions around how and why confirmatory expectations can generate both generate productive and problematic behavior. 
Through a moment-by-moment analysis of these students' framing and the dynamics of their behavior, we seek to understand what constitutes productive confirmation framing, as part of understanding more broadly why, how, and when students take up opportunities to engage productively in scientific practices.

We begin in section \ref{background} with prior research on learning and instruction for physics labs, outlining the development of "reformed" lab environments and then focusing specifically on confirmatory expectations and student behavior in labs. 
Much of that work, and ours here, draws on theoretical constructs of \textit{framing}, \textit{epistemic agency}, and \textit{disciplinary practices}.
In section \ref{theory}, we return to expand and clarify the literature's and our use of these constructs. 
We outline our data collection and production methods in section \ref{Methods} before turning to the data and analysis of the episode in section \ref{Data}, to argue that it reflected students' productive engagement in empirical science that was, in part, driven by their framing the activity as confirming a result.  
Finally, we discuss implications for future research and lab instruction in sections \ref{discussion} and \ref{Implications}. 

\section{Research and development in introductory physics labs}
\label{background}

\subsection{Research-Based Lab Pedagogy}
Physicists, educators, and education researchers have long expressed dissatisfaction with traditional laboratory instruction (e.g. \cite{reif_teaching_1979, thornton_tools_1987, otero_past_2017, smith_best_2021}), which we take as defined by \textit{i)} aims to reinforce content knowledge \textit{ii)} activities that demonstrate or verify physical concepts \textbf{and} \textit{iii)} well-defined, prescriptive experimental protocols for students to follow. 
Research consistently indicates these learning environments are ineffective:  they do not measurably contribute to students' conceptual understanding \cite{etkina_design_2010, holmes_value_2017, smith_direct_2020} and they negatively impact the development of expertlike epistemological attitudes and beliefs \cite{wilcox_open-ended_2016, wilcox_developing_2017, hu_qualitative_2017, smith_direct_2020, walsh_skills-focused_2022}.

In 2014 the American Association of Physics Teachers (AAPT) produced a framework of learning goals for labs, with 6 interconnected foci:  constructing knowledge, modeling, designing experiments, developing technical and practical laboratory skills, analyzing and visualizing data, and communicating physics \cite{AAPT}.
As Smith and Holmes \cite{smith_best_2021} argue, rather than changing lab instruction to better meet the goal of reinforcing content covered in lecture, labs should focus on teaching students what it means to do experimental physics – the skills and activities encapsulated by the AAPT lab goals foci. 

A number of groups have been working on reforms for introductory undergraduate labs including the Investigative Science Learning Environment (ISLE, \cite{etkina_design_2010}); Thinking Critically in Physics Labs, first at the University of British Columbia \footnote{\url{https://www.physport.org/curricula/thinkingcritically/}} \cite{holmes_quantitative_2015, holmes_teaching_2015} and continuing at Cornell \cite{holmes_operationalizing_2019, kalender_restructuring_2021, phillips_not_2021, sundstrom_instructing_2023}; and the Design, Analysis, Tools, and Apprenticeship lab (DATA lab, \cite{funkhouser_design_2019}). 
There have been efforts in other disciplines as well, including, in biology, hybrid computational-experimental labs \cite{gouvea_motivating_2022}, the Authentic Inquiry through Modeling in Bio labs (AIM-Bio; \cite{hester_authentic_2018}), and Course-based Undergraduate Research Experiences \cite{auchincloss_assessment_2014}.
These curricula share theoretical motivations and goals of supporting student epistemic agency and meaningful disciplinary engagement, but they take different approaches.

ISLE, the DATA lab, and the first versions of the Thinking Critically in Physics labs focus on identifying and scaffolding component skills and scientific abilities.
These labs seek to make the critical thinking process explicit to students, to demonstrate its effectiveness and also to provide support "in a deliberate way with targeted feedback" (\cite{holmes_teaching_2015}, p. 1). 
In the ISLE and DATA labs, the rubrics that underpin their formative assessment operate in a similar fashion: making explicit the component skills for expert-like behavior so that students can align their behavior with experts.  

Gouvea \textit{et al.} \cite{gouvea_motivating_2022} and Hester \textit{et al.} \cite{hester_authentic_2018} take a different approach in their biology labs. 
Rather than identifying the specific forms of disciplinary activity they are guiding students towards with explicit scaffolds, they develop contexts that are authentically complex and uncertain, where meaningful scientific practices are likely to emerge. 
This includes focusing on phenomena with multiple possible opportunities for empirical discrepancies, and reconfiguring assessment structures to focus attention on student ideas and reasoning instructors. 
They expect students have productive knowledge and abilities, position them with significant epistemic agency, and ask of instructors to recognize and cultivate the beginnings of scientific inquiry that emerge. 
Recent versions of the Thinking Critically in Physics labs work to provide contexts for student agency, in activities without pre-determined correct outcomes, to encourage revision, iteration, and exploration (e.g., \cite{sundstrom_instructing_2023}. 

Labs that support student agency improve epistemological attitudes and beliefs \cite{brewe_modeling_2009, smith_direct_2020, wilcox_developing_2017, dounas-frazer_student_2017} and support productive experimental activity \cite{etkina_design_2010, sundstrom_problematizing_2020, descamps_case_2022, sundstrom_instructing_2023}.
Students in such "reformed" labs engage more in sensemaking and experimentation \cite{etkina_design_2010, smith_direct_2020}, more in both quantity and quality than students in traditional lab contexts \cite{nixon_undergraduate_2016, holmes_developing_2020}. 

The evidence is sufficient, we believe, to warrant a general shift away from traditional labs. 
The work from here is to develop and refine these approaches.
Researchers assert the benefits of various features – complex, open-ended activities; opportunities for iteration; multiple modalities of investigation and representation; and student responsibility for the experimental design \cite{etkina_design_2010, hester_authentic_2018, holmes_developing_2020, gouvea_motivating_2022}.
This is by no means a comprehensive list, and how these features interact with other parts of the curricular system is non-trivial \cite{Zwickl_characterizing_2023}.

One clear reason to continue research on the dynamics of student inquiry in these labs is that students do not all take up opportunities to engage in scientific inquiry \cite{phillips_not_2021}.
Part of the challenge seems to be in their expectations, or, more specifically, \textit{framing}, regarding what happens in instructional labs \cite{sundstrom_instructing_2023}. 
In particular, students tend to frame labs as exercises in confirmation, expecting that successful performance constitutes their arriving at an already-known correct answer. We turn to that literature now.

\subsection{Confirmation framing in instructional labs}
Students' experiences of discrepancies can lead to sensemaking \cite{apedoe_empirical_2010, may_student_2022} but they do not always \cite{phillips_not_2021}. 
For example, in one lab from the Thinking Critically in Physics curriculum students measure the acceleration of falling objects and compare the results to the predictions of two models, one free fall and the other including drag (air resistance). 
When the falling object is a beach ball, neither model works.
Some students notice the discrepancy and engage with it as a problem \cite{sundstrom_problematizing_2020}, but others do not \cite{phillips_not_2021}. 
Phillips \textit{et al.} \cite{phillips_not_2021} highlight that the presence of opportunities to construct scientific knowledge is only one component to productive behavior: students also need to perceive these opportunities (see also \cite{miller_addressing_2018, hayes_destabilizing_2020, smith_how_2020}). 

What students focus on, react to, and consider appropriate action in an instructional laboratory is connected to their expectations (e.g., \cite{scherr_student_2009}). Students often enter instructional laboratories expecting to follow prescriptive procedural instructions and to be evaluated on the correctness of their findings \cite{hu_qualitative_2017, smith_how_2020} . 
In Wilcox and Lewandowski \cite{wilcox_students_2017}'s analysis of a large-scale survey of views towards experimental physics, 66\% of students in first year courses agreed with the statement, "The primary purpose of doing physics experiments is to confirm previously known results." 
In a follow up study examining students' explanations and justifications for their non-expertlike responses, Hu \textit{et al.} \cite{hu_qualitative_2017} found many students discussed how experiments in the classroom context are supposed to support their conceptual learning – by demonstrating concepts and reaffirming what is already known. 

In additional qualitative analyses of students' epistemological views, Hu and Zwickl \cite{Hu_Zwickl_2017, hu_examining_2018-1} found that 93\% of introductory students consider the role of labs to primarily be supplemental learning experiences for theories and concepts. 
Furthermore, they found that about 46\% of students considered agreement with theoretical predication a suitable basis for establishing the validity of experiments. 
Given that these students also rarely recognized uncertainty analysis as a tool for establishing validity, as Hu and Zwickl \cite{hu_examining_2018-1} discuss, these views may lead to a distorted understanding of the nature of science. 

Indeed, Smith \textit{et al.} \cite{smith_how_2020} and Phillips \textit{et al.} \cite{phillips_not_2021} document how student expectations to confirm concepts from lecture inhibit productive behavior in lab and engender questionable research practices. 
Smith \textit{et al.} \cite{smith_how_2020} observe students manipulating their experiment and data analysis in order to match theoretical predictions. 
Phillips \textit{et al.} \cite{phillips_not_2021} describe students as treating the activity as "a series of hoops to jump through to fulfill assignment requirements," which, they argued, kept them from problematizing the discrepancy.

The students surveyed in Smith \textit{et al.} \cite{smith_how_2020} expressed similar expectations about the purpose of their instructional lab as enhancing or supplementing lecture content. 
Semi-structured interviews with a set of these students suggests that prior and concurrent experiences in both high school and non-physics college labs contribute to this framing. 
Even as lab designs aim to prioritize experimentation, shifting how students perceive and approach learning opportunities is not easily accomplished \cite{hayes_destabilizing_2020, phillips_not_2021, gouvea_motivating_2022, sundstrom_instructing_2023}.

Hayes and Gouvea \cite{hayes_destabilizing_2020} similarly document a student, Caleigh, understanding lab as "about demonstrating a target idea" in contrast to the curricular aims. 
In an interview, this student explained that because all lab groups were given the same experimental procedures, she expected deviations or discrepancies meant mistakes. 
Sundstrom \textit{et al.} \cite{sundstrom_instructing_2023} describe explicit efforts made by a graduate lab instructor to shift students from a focus on confirmation to one on falsification, which were mostly (but not always) successful in changing behavior and supporting productive behavior. 
Hayes and Gouvea \cite{hayes_destabilizing_2020} also highlight how Caleigh made progress over the semester, coming to take up epistemic agency in activities that had students devise their own experiments, toward genuinely open-ended results. 

Evidence across these studies suggests that students' expectations that lab are about labs confirming a known result – confirmation framing – can remain stable and can inhibit their recognizing and engaging with discrepancies. 
Our analysis below concerns an episode in which, we argue, confirmation framing supports students' engagement with discrepancy. 
Before we proceed to it, we pause to fill in theoretical background on the constructs of \textit{framing}, \textit{epistemic agency}, and \textit{disciplinary practices}.

\section{Theoretical Background}
\label{theory}

Our purpose in this section is to clarify our particular use of these terms, given that they may be novel for some readers and that they have been used with different nuances of meaning. 
Overall, we understand these constructs to conceptualize aspects of highly complex dynamics that involve multiple scales of time and system \cite{suarez_learning_2023, thelen_smith}.

\subsection{Framing}
A frame or framing is an activated or emergent "structure of expectations" \cite{tannen1993framing} that reflects and influences how people understand a situation, what could happen, what features require attention, and what actions are appropriate. 
The construct's uses across disciplines reflect a wide range of scales. Linguists Tannen and Wallat \cite{tannen_interactive_1987} studied framing on the scale of interactions in a small group during a medical examination; Goffman \cite{goffman1974} focused on how a society shapes and is shaped by ways its members share framing situations. 
In our analyses, we examine how behavioral, verbal, and paraverbal cues communicate metamessages about an individuals' framing, influencing others' framing and regulating a groups' activity. 
When relevant, we also speak of the groups' framing and consider the behavioral, verbal, and paraverbal metamessages as reflecting the group as a whole \cite{scherr_student_2009}. 

Smith \textit{et al.} \cite{smith_how_2020}, Hayes and Gouvea \cite{hayes_destabilizing_2020}, Phillips \textit{et al.}  \cite{phillips_not_2021}, and Sundstrom \textit{et al.} \cite{sundstrom_instructing_2023} demonstrate how students' framing shapes their engagement. 
Students bring a wealth of past experience into their classroom that they use, explicitly and tacitly, to organize and interpret what is taking place. 
Cues from instructors, peers, and instructions – intended or unintended – have significant consequences for the nature and quality of participation in such learning environments. 
Even though the lab activities in Smith \textit{et al.}  \cite{smith_how_2020}, Hayes and Gouvea \cite{hayes_destabilizing_2020}, and Phillips \textit{et al.}  \cite{phillips_not_2021} intended for the students to construct knowledge, some students still framed the activity as confirming already known information. 

Furthermore, we interpret the findings of Wilcox and Lewandowski \cite{wilcox_students_2017} and Hu and Zwickl \cite{Hu_Zwickl_2017, hu_examining_2018-1} as evidence for confirmation framing being a stable and regular phenomenon among undergraduate students, reflecting patterns in their experiences over years in schools. 
The students surveyed in Smith \textit{et al.} \cite{smith_how_2020} expressed similar expectations about the purpose of their instructional lab. Semi-structured interviews with a set of these students suggests that prior and concurrent experiences in both high school and non-physics college labs contribute to this framing. 

Jim\'{e}nez-Aleixandre \textit{et al.} \cite{jimenez-aleixandre_doing_2000} describe another problematic framing, students' seeing the work in science class as completing tasks they are assigned rather than about working to build understanding, "doing the lesson" rather than "doing science." 
Phillips \textit{et al.} \cite{phillips_not_2021} call this framing "hoops," as in 'jumping through hoops,' and argue that it can help explain students' not problematizing conflicts between their data and their understandings. 
Gouvea \textit{et al.} \cite{gouvea_motivating_2022} discuss how prescriptive laboratory instructions and assessment rubrics can cue a 'checklist' approach to writing \cite{tang_reconsidering_2015}, and so shift their assessment structures to emphasize sensemaking. 
Across these accounts is a concern for students not participating in the process of knowledge construction, that is having \textit{epistemic agency} We turn now to a discussion of epistemic agency.

\subsection{Epistemic Agency}
In our work, we focus on epistemic agency to organize our exploration of what constitutes productive participation in scientific knowledge construction. 
Epistemic agency is the prerogative for individuals or groups to participate in "the whole range of components of knowledge building-goals, strategies, resources, evaluation of results, and so on" (\cite{Scardamalia_Bereiter_2005}, p. 108). 
Enabling and supporting epistemic agency is a common goal of contemporary introductory lab curricula \cite{hester_authentic_2018, hayes_destabilizing_2020, holmes_developing_2020, kalender_restructuring_2021} and many other science learning environments \cite{stroupe_examining_2014, mcpadden_productive_2020, manz_rethinking_2020, phillips_physicality_2023}.

With regards to curricular design and instruction, scholars have long argued that research and instruction should focus on the forms of epistemic agency available and valued in particular learning environments \cite{gresalfi_constructing_2009, stroupe_examining_2014, miller_addressing_2018, stroupe_introduction_2019}. 
This perspective highlights the role of perception and social dynamics in students' recognition of and engagement with opportunities to build knowledge. 
Indeed, studies of students framing in introductory labs have attended specifically to how students' framing what is taking place in labs affords or limits their epistemic agency \cite{hayes_destabilizing_2020, sundstrom_problematizing_2020, phillips_not_2021,sundstrom_instructing_2023}.

As mentioned above, Hayes and Gouvea \cite{hayes_destabilizing_2020} highlight how as Caleigh's framing shifted away from "demonstrating a target idea" – a change the materials and instructors worked to promote – she perceived more opportunities to participate in knowledge building. 
Sundstrom \textit{et al.} \cite{sundstrom_instructing_2023}, in contrast, described students' limited progress despite a lab instructor's efforts specifically to challenge confirmation framing. 
While the TA's efforts were largely successful at shifting students' framing away from confirmation, that did not necessarily result in their greater epistemic agency. 
These studies motivate our exploration of when, how, and why students enact epistemic agency in introductory labs.

For our analysis here, we draw from Dam\c{s}a \textit{et al.} \cite{damsa_shared_2010} to identify evidence of epistemic agency. 
They define "shared epistemic agency" to describe the process by which groups work together to deliberately produce knowledge and identify evidence of agency at the scale of a collaborative group. 
Their framework focuses on the actions that guide and influence the activity. 
Specifically, they consider two main dimensions: actions that contribute to the development and refinement a knowledge object (epistemic) and actions that organize the process of knowledge creation (regulative). 
Epistemic actions are productional in nature – actions that contribute to the production of ideas and artifacts. 
Regulative actions encompass management the epistemic-productional work, from goal setting to the coordination of activities and the social negotiation necessary for collaboration.

\subsection{Discplinary Practice}
Central to our understanding of productive behavior in learning environments is the long-standing goal of science education research and reform to engage students in "doing science" themselves \cite{NGSS, hofstein_laboratory_2004, duschl_science_2008, AAPT, otero_past_2017}.
The goal, or part of it, is that students' epistemic agency is directed towards "the pursuit of coherent, mechanistic accounts of natural phenomena" (\cite{hammer_identifying_2008}, p. 13).

We adopt a science-as-practice view in our conceptualization disciplinary activity \cite{lehrer_scientific_2007, ford_chapter_2006, stroupe_examining_2014, manz_rethinking_2020}.
We draw from Pickering \cite{pickering_mangle_1995} and Ford \cite{ford_educational_2015} to connect learners' enactment of epistemic agency to disciplinary practice. 
In Pickering's \cite{pickering_mangle_1995} account, scientists develop ideas, measures, and apparatuses, which they apply to the material world, and the world responds, often to "resist" scientists' efforts. 
They respond in return, modifying their procedures, understandings and hypotheses, or even shifting their aims. 
This "dialectic of resistance and accommodation" (p. 22) is a complex and cyclic interaction that, hopefully, progresses toward alignment, which Pickering emphasizes is an ill-defined goal: scientists cannot know in advance what it will take to capture and understand a phenomenon. 

This sense of emergence parallels the argument that Ford \cite{ford_educational_2015} forwards for the affordances of choosing "practice" to describe science. 
Rather than a timeless and absolute set of rules, a determinate definition of how science works, talking about science-as-practice (or as comprised of diverse "practices") focuses our attention on how the doing of science forms, emerges, and shifts. 
As Ford \cite{ford_educational_2015} puts it, the normative "correctness" of a component performance is defined by how it interacts with other performances in service of disciplinary aims and goals. 
Put simply, actors have many resources at their disposal – procedures, machines, other actors, concepts, hypothesis, representations – and the appropriateness of ones' use of those resources depends on how it interacts with and responds to other activity and material world. 

Furthermore, actors have purposes in their activities. 
Yet, as Pickering \cite{pickering_mangle_1995} emphasizes, these purposes are themselves liable for revision in response to shifting understandings. 
Ford \cite{ford_educational_2015}, too, discusses the prospective nature of disciplinary aims and the implications for practice: "Evaluations of performances and the interactions of performances are never definitive but tentative" (p. 1045). 
The substance of what scientists do and to what end they enact agency shifts, but the interactions and processes that stabilize patterns of activity remain coherent over time. 

A semantic, but nevertheless important point to clarify is the difference between practice and practices: the enterprise of science (scientific practice) is comprised of locally constructed regularities in activity (scientific practices). 
We chose not to take up Ford's \cite{ford_educational_2015} terminology of "performances" but still conceptualize scientific practices as the "constituent activities" of science (p. 1043; see also \cite{manz_rethinking_2020}). 
Our use of "practices" – scientific, disciplinary, experimental – not only follows other researchers (e.g. \cite{berland_epistemologies_2016, phillips_beyond_2018,  brewe_modelling_2018, manz_rethinking_2020,gouvea_motivating_2022, Zwickl_characterizing_2023} but also makes clear the connections to national recommendations (e.g., \cite{NGSS, AAPT}).

\section{Data and Methodology}
\label{Methods}

\subsection{Instructional Context}
This case study comes from an introductory physics lab intended to promote student autonomy in designing experimental methods and drawing their own conclusions. 
There were four lab activities in the course, each lasting 2 or 3 weeks. 
We examine work here by students Holly, Judy, and Peter (all pseudonyms) during the second and third weeks of the first lab in Spring 2021, when the lab was operating in a hybrid format due to COVID-19. 
Students worked in groups of two or three, with some students in-person (Holly and Peter, in this group) and others virtual (Judy).

This case study primarily comes from the students' work in the second week of lab, when students were still conducting experiments, analyzing their data, and generating conclusions. 
The third week of this lab activity was entirely devoted to group presentations and a reflective discussion on their first lab activity. 
We also include Holly, Judy, and Peter's presentation as part of this case study.

The lab assignment focused on a simple pendulum. 
It described Galileo's claims that the period of does not depend on either the mass of the bob or on the amplitude of the swing. 
During the first week of lab, as implemented in this semester, the assignment was to "Design and carry out an experiment to investigate and test the predictions of Galileo's model." 
During the second week, the instructions encouraged students to "improve the precision" of their measurements, with the assignment to "determine quantitatively" the precision of their measurements, and to decide – providing an "unambiguous statement" – whether their "results are consistent with Galileo's model." 
Students could choose whether to work on mass or amplitude; Holly, Judy, and Peter worked on amplitude. 

In fact, the period does depend on amplitude, due to the divergence of a pendulum from simple harmonic motion, which careful measurement can show. 
In this respect, the lab provides the opportunity for students to encounter a discrepancy between Galileo's claim and their measurements. 

\subsection{Data Collection and Methodology}

With the hybrid modality of labs, we collected data by having students record their group video calls and the TA record the virtual whole class discussion. 
We also had a camera and audio recorder in the lab room.

The first author watched the recorded video calls, logging student activity in 5-minute intervals and taking notes. 
Then they selected candidate episodes for analysis, based on prior literature on sensemaking and scientific inquiry; indicators included mechanistic reasoning \cite{russ_recognizing_2008, russ_making_2009} vexation or problematizing \cite{phillips_beyond_2018, odden_vexing_2019, odden_defining_2019}, not-understanding or uncertainty \cite{watkins_positioning_2018, jordan_managing_2014}, and argumentation \cite{driver_establishing_2000, ford_dialogic_2012}. 

Analysis of episodes followed methods described in Hammer \textit{et al.} \cite{hammer_idiosyncratic_2018}, with the first author in this case taking the lead. 
They prepared an analytic memo \cite{bailey_guide_2022}, a draft narrative account of what took place citing evidence to support interpretations, as well as marking interpretive uncertainties. 
They then presented this draft to the research team in a series of sessions. 
The team worked from the draft, watching and rewatching each segment of the video, accepting,  revising, and challenging claims within it. 
The first author revised the document based on discussion and feedback, to arrive at the consensus narrative analysis we present below. 

We selected this episode for this paper, because it constituted a kind of discrepancy itself: The students in it seemed to be framing lab as confirmation, but they seemed to be engaged in doing science. 
We turn now to present and analyze that data.

\section{An episode of productive confirmation framing}
\label{Data}

Holly, Judy, and Peter produced data for amplitudes of 10\degree, 20\degree, 30\degree and 40\degree by timing five swings of their pendulum and dividing by five to find the average period; they did five trials at each amplitude. They estimate their uncertainty in timing to be $\pm0.2$ seconds, which makes the uncertainty in each period measurement $\pm0.04$ seconds. To show their data, we provide as Fig. 1 and fig 2, two graphs the students produce toward the end of the lab period but note they had not produced this graph at this point in the episode. 

\begin{figure}
  \includegraphics[width=1\linewidth]{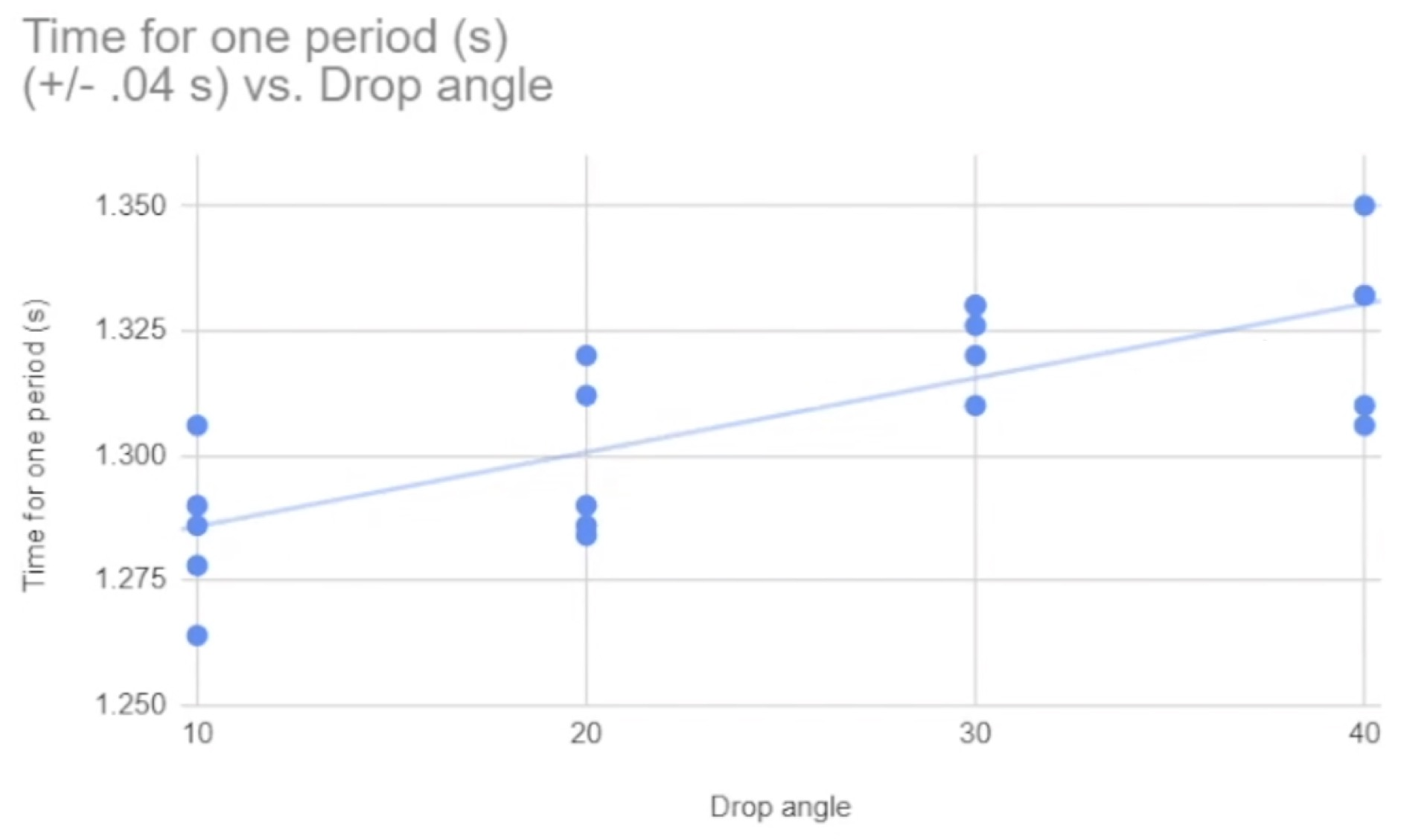}
 \caption{Students' pendulum data. A scatter plot showing all the five trials at each amplitude. \label{fig1}}
\end{figure}

\begin{figure}
  \includegraphics[width=1\linewidth]{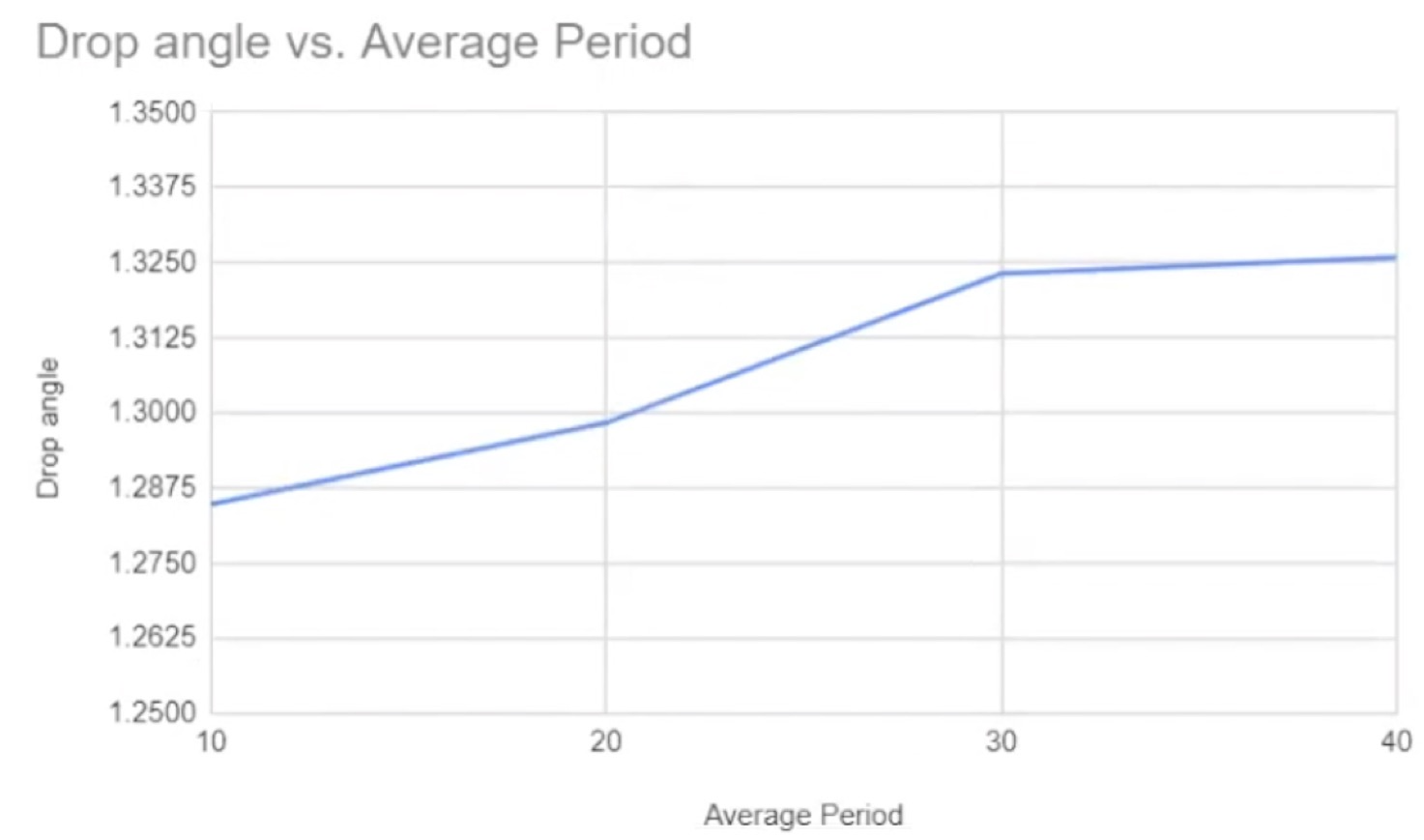}
 \caption{Students' pendulum data. A line plot showing the average period length for each amplitude \label{fig2}}
\end{figure}

We present the data and analysis in four segments:
\begin{enumerate}
  \item Confirmation Framing and Problematizing
  \item Productive Epistemological Framing:  Troubleshooting
  \item Productive Epistemological Framing:  Data Representation
  \item Questionable Research Practices
\end{enumerate}

\subsection{Confirmation Framing and Problematizing}
This episode begins immediately after they finish collecting data, a little over an hour into their lab period. 

\begin{displayquote}
\textbf{1. Peter:} I mean so we are clearly seeing like very slight changes. I know, I think we had the same thing last time that the amplitude seemed to change it just a tiny bit. I wonder what about like how we're doing it is making it change it consistently?

\textbf{2. Holly:} Is it not supposed to change?

\textbf{3. Peter:} 
Uh, it should be the same regardless of amplitude. 
But I guess there's, there must be something else that we're doing that's making it change just a little bit. 
Although it's very insignificant.

\textbf{4. Holly:} I wonder if it's, um, friction. Like, is it-- does this move back and forth [reaches up to examine pendulum string]-- No
\end{displayquote} 

Peter sees the data as "clearly" showing "slight changes" in the period with amplitude, which he remembers their seeing in the first lab session as well. 
He locates the problem in their doing of the experiment: something they did made the data change in this way. 
With Holly's question about what is supposed to happen and her suggestions in line 4, we infer that both Peter and Holly are framing their lab activity as confirming a known, correct result. 
Judy, on the other hand, suggests a different interpretation of the data:

\begin{displayquote}
\textbf{8. Judy:} But I feel like the correlation is too strong to ignore, like it-- like it makes sense, like it's decreasing very slightly as you decrease the amplitude

\textbf{9. Peter:} Yeah, I mean, I think looking at our data, it would seem that it is related, but just at a really small like ratio I guess, so like the, uh, the amplitude has a really small effect. But like, I know, theoretically we shouldn't be seeing any effect. So I'm wondering what about what we're doing is making it look like that.

\textbf{10. Holly:} I don't know. Like, yeah, like where is our error coming from?
\end{displayquote} 

In line 8, Judy's saying "but" suggests she distinguishes her interpretation from Peter's and Holly's, that the effect they see in their data could be a feature of the phenomenon.
Her reasoning invokes a mathematical intuition that the change is so consistent and clean, as opposed to noisy, that it makes more sense that these data reflect a physical phenomenon rather than some experimental error. 
She does not seem to frame the goal of their work as confirming Galileo.

Judy makes a bid for their seeing the data as reflecting a physical phenomenon, without invoking notions of error or wrongness. 
In part, her bid is for accountability to the data. 
We argue that this move contributes to an expectation that their explanation of their data, whether for or against Galileo's model, is accountable to the data.

In line 9, Peter acknowledges the logic of Judy's reasoning, that based on their data "it would seem" the period changes. 
Still, he maintains there should be no effect, vesting epistemic authority in Galileo (or perhaps the instructor). 
Peter and Holly frame the problem as finding "what about what we're doing is making it look like that" (line 9), "where is our error coming from" (line 10). 
The problematizing evident in these lines is a demonstration of epistemic agency: the students identify a discrepancy between their data and the theoretical model and then, in dialogue with each other, they refine their understanding of what is the problem to solve. 

Holly and Peter then begin to brainstorm ideas for where their error could come from. 
Judy, however, does not participate. 
She is remote, which likely plays a significant role in her lack of participation in that activity, and her interest seemingly lies in closer examination of the data. 
After Peter remarks "Although it's very consistent for some reason" (line 13) Judy suggests they graph their data and starts working on that. 

\subsection{Productive Epistemological Framing:  Troubleshooting}
While Judy works silently on processing their data, Holly and Peter turn to their apparatus to talk through error and uncertainty in their procedure. 

\begin{displayquote}
\textbf{19. Holly:} I wonder if error could also be in the drop itself. Like if you don't just like [random noises] take your hand directly away, like if it's like cushioning it at all. But like, I don't know how that would...

\textbf{20.	Peter:} Yeah and, I mean, would that have a larger effect at higher drop height? That's the question.

\textbf{21. Holly:} I don't know. I mean actually it might
\end{displayquote} 
Exploring whether this idea might work, Peter brings up an important consideration of their data:
\begin{displayquote}
\textbf{24. Peter:} Yeah, so I mean, that could be it. Do we think, can we think of a reason why it would be a larger effect at a higher drop. I mean, say it's staying exactly where it is for a little bit before it starts to fall

\textbf{25. Holly:} Um, well I don't know. Cause like if you only bring it out to here it has more velocity in the x direction than the y direction whereas up here [pendulum set up starts to fall over]-- [...] If you bring it up here, like it has like, past-- well it's not like we're going passed 45 degrees, but like there's more velocity in the y direction up, up at a higher point... maybe
\end{displayquote} 

Holly suggests a possible physical effect in releasing the bob that might affect the pendulum's motion; Peter asks if that effect would vary with amplitude. 
They frame their activity as looking for physical mechanisms to explain the trend in their data. 
It is possible Judy's critique not to ignore the correlation had an influence: The physical, mechanistic explanation they are seeking must be accountable to the data and, perhaps, to Judy as well.

Their first ideas, that the error might come from timing or from the drop, are not sufficient explanations to them. 
Rather, they are the beginning of a search for a mechanism that would consistently increase the period length as they increase amplitude. 
Their attention is clearly focused on the apparatus within the emergent reasoning context of figuring out their data. 

We claim this work is disciplinary troubleshooting as Smith \textit{et al.} \cite{smith_how_2020} anticipated and in line with Pickering's account of a "dialectic of resistance and accommodation" (\cite{pickering_mangle_1995} p. 22). 
Data from their apparatus "resists" their expectations, and they are working to "accommodate," to figure out how to modify their apparatus. 
Their expectations, meanwhile, reflect their confirmation framing of the lab as a whole. 
That is, they expect to confirm the authoritative claim, but the apparatus resists with discrepant data. 
Thus, we claim, for Holly and Peter in this moment, confirmation framing supports disciplinary engagement with discrepancy. 

\subsection{Productive Epistemological Framing:  Data Representation}
Judy, who has been engaged in her own activity – separate from Holly and Peter both by her remote participation and by her different framing – returns to the conversation a few turns of talk after line 25. 
She has been working on a spreadsheet, to help study the data. They look at a graph of the data that they say is a straight line and which turns out to be an incorrect representation. 
(They did not save the graph, and the video does not show it.)  

\begin{displayquote}
\textbf{40. Judy:} Cause, yeah um, it looks like it's really linear.

\textbf{41. Holly:} Yeah.

\textbf{42. Judy:} Do you see this? [laughs]

\textbf{43. Peter:} Yeah. Wha-- uh...

\textbf{44. Holly:} Guys we just disproved Galileo's theory

\textbf{45. Judy:} It's a literal straight line!

\textbf{46. Peter:} No, no, no. It's, it-- this is not a straight line for the reason you think this is a straight line, look at the axes flipped [laugh]

\textbf{47. Judy:} But like it's so straight, like the line is like

\textbf{48. Peter:} I mean the reason it's straight is because we're, this isn't an actual x-axis on the bottom, like look at what the units are on the bottom

\textbf{49. Judy:} Oh, I'm so silly [laughs]
\end{displayquote}

Judy's sharing her spreadsheet prompts Peter and Holly to shift their attention back to the data. 
Looking at the clear, "straight line" of the graph, Holly and Judy conclude that they have disproved Galileo, that amplitude affects period. 
Holly, who was previously invested in figuring out how their experimental procedure caused error, now exclaims "we just disproved Galileo" (line 45). 
Her tone sounds joking, perhaps because she thinks the idea is absurd, or perhaps as "epistemic distancing" \cite{conlin_making_2018}. 
Judy, on the other hand, is clearly happy about this development.

Peter, however, notices a mistake. 
We do not have direct evidence of the mistake, but we reproduced what we suspect it was: they included the word "seconds" in the cells with their raw data, and the period values are to the left of the amplitudes. 
This would have Excel ignore the period values and plot (1, 10), (2, 20), (3, 30), and (4, 40) – a straight line that does not reflect their data. 

Judy recognizes the mistake and expresses responsibility for it – "I'm so silly." 
In the work that follows she steps aside, asking, "Can somebody else attempt this, I'm, maybe I'm just terrible at Excel but it's literally just not letting me" (line 59). 
She participates less and less as the session continues, perhaps feeling a loss of self-efficacy or legitimacy after troubles with Excel. 
It had already been a challenge for her working remotely through Zoom. 

The group continues in its reexamination of the data, a shift from Holly's and Peter's framing of looking for physical mechanisms. 
We return to the students' work as they examine a new graph, which again we cannot see. 
Holly leans towards her computer and says,

\begin{displayquote}
\textbf{64. Holly:} It almost looks like it dips down at 30

\textbf{65. Judy:} Yeah, it could just be from like error. But they're all like 1.3 something, {like they're very close--}

\textbf{66. Peter:} {Oh. That's, that's that's} the issue, it's cause this was, uh, was 1.28 here and 1.33 and [Unclear] uh cause we had it 30 20 10 40
\end{displayquote}

Here we see Judy shift in her approach to the data and to the activity. 
Even though her comment in line 65 is an evidence-backed explanation, she now seems to view the data as relatively consistent. 
That she says "just" seems to indicate that the abstract notion of "error" is a now sufficient explanation for the data, clearly a shift in her thinking. 

Peter's comment in line 66 suggests that the current plot has the amplitude data incorrectly ordered. 
As Figure 2 makes clear, reordering the amplitude data would undercut the evidence for a linear relationship, possibly influencing Judy's new explanation as well as her reduced level of participation. 
It is also reasonable to suspect she found the exchanges in lines 40-50 dispiriting, but there is limited evidence. 

Following line 66, Peter and Judy briefly discuss some spreadsheet logistics. 
Peter narrates some data reformatting as he works to generate what seems to be an accurate graphical representation of their data, Figure 1. 
Holly examines that graph.

\begin{displayquote}
\textbf{69. Holly:} It's interesting how much closer the 40-degree and 30-degree values are compared to the 10-degree and 20-degree ones

\textbf{70. Peter:} Yeah

\textbf{71. Holly:} So I wonder if using a bigger amplitude would have us-- or, a larger amplitude would allow us to have more accurate results in accordance with the theory

\textbf{72. Peter:} It's possible. Yeah I think it would probably be easier to measure for larger amplitude because we are, you know, it's a more extreme peak.

\textbf{73. Holly:} So then maybe it is a timing error for the smaller ones.

\textbf{74. Peter:} It might be. 
\end{displayquote}

In addition to her comments in lines 64 and 68, Holly's leaning towards her computer is evidence she is focused on interpreting the data. 
Her observation and conjecture in lines 69 and 71 reflect and support a return to their earlier efforts, as she considers what the data suggests might be a mechanism for how their experiment shows a dependence. 
Peter joins this work, articulating this explanation in his own words and elaborating on the mechanism. 
Holly and Peter are thus focused again on how they produced the data, with new ideas from new insight from the graph into what it shows. 

For the next 18 turns of talk, Holly and Peter discuss the formatting of different representations of their data, plotting it in different ways with Excel. 
To use the language of knowledge-creation, Peter and Holly are invested in generating knowledge-objects to concretize their ideas; this Excel reformatting is a demonstration of shared epistemic agency. 
They are attending to key procedural aspects of their experiment and taking these procedural actions seriously. 

Notably, they engage in this behavior while still framing the activity as confirming a known result. 
In addition to problematizing, troubleshooting, sensemaking observed earlier, their confirmation framing leads them to produce accurate graphs, engage in graphical analysis, and data interpretation. 
Across this extended work, they retain epistemic agency, organized by their (epistemic) framing of the activity. 

Judy, meanwhile, participates less and less. 
She tries to interject at one point, but Peter talks at the same time, and neither he nor Holly seem to recognize that Judy had started to speak.

\subsection{Questionable Research Practices}
The students have Excel put in a generic trendline for their data, which spurs the following exchange:

\begin{displayquote}
\textbf{92. Holly:} Unfortunately it's not super horizontal

\textbf{93. Peter:} Yeah, I mean it, I think actually it is close enough to horizontal because of how small our axes is. Like if I zoom this out, uh, if I go from like 0 to 2

\textbf{94. Holly:} Gotcha

\textbf{95. Peter:} Like it's extremely horizontal

\textbf{96. Holly:} Do we want to like make another copy of this graph, show a zoomed in version versus a zoomed out version?

\textbf{97. Peter:} We could do something like that.

\textbf{98. Holly:} And be like, despite what it looks like this line is actually fairly horizontal.

\end{displayquote} 
Line 92 is a representative description of how Holly and Peter have been interpreting their data: "Unfortunately, it's not super horizontal." 
Holly is honest about what their graphical analysis demonstrates and understands it to be a problem. 
While Peter had previously shared in this interpretation of their data (starting at line 1), his response in line 93 seems to be a departure: "it is close enough to horizontal," so it is actually not a problem. 

Rescaling plots to gauge the relative size of an effect is a reasonable, if novice, analytic technique. 
Still, Peter's explanation is not based in physical or mechanistic reasoning – that the graph looks horizontal is the evidence, and he makes no mention of the uncertainties in the group's data. 
At the same time, it does not seem to us that Peter means to manipulate their data to hide this problem. 
In line 94, Holly signals her agreement by suggesting that they show both graphs. 
Holly has her own conditions for being satisfied with this explanation and conclusion to their inquiry: intellectual honesty and epistemic accountability. 

Still, it is here we see the first evidence of confirmation framing leading to questionable research practices. 
The group saw the data showing a small variation in period with amplitude, in this lab session as in the previous session, but with their strong expectation that there should be no variation, they favored the representation that minimized it. 
While their graphical analysis is questionable and exaggerates their uncertainty\footnote{To be fair, their data does not clearly contradict Galileo's model:  the difference between their average measured period at 10\degree and 40\degree is 0.042 seconds with an uncertainty of $\pm$ 0.04 seconds.}, they seem to believe their final explanation is appropriate. 

We stop the transcript of the main episode here, but a few turns later Peter asks Judy for her thoughts on the new zoomed out graph (line 97), clearly a bid for her participation, and perhaps a bid for consensus. 
She responds that she cannot see the graph. 
After the group works it out for her to see, she seems to agree, saying "Oh okay yeah yeah" (line 107) and "Yeah I think we should include both [graphs] then" (line 109). 
Shortly later, the TA enters their video call, which changes the activity and ends the episode.
It seems unlikely that Judy would have changed her mind, but we cannot rule out that, if the TA had not joined at that moment, Judy would have continued this conversation and engaged with the conceptual substance of these graphs. 

The following week the students presented their conclusions to their classmates. 
The lab TA asked each group to select one plot or image and encouraged students to ask each other questions. 
For their turn, the students showed Figure 1 and a zoomed-out version of that graph. 
Holly started the presentation by summarizing how they determined their uncertainty. 
Then Peter and Judy presented their data and conclusion.

\begin{displayquote}
\textbf{P1. Peter:} The one thing we put on this slideshow, we actually put two plots. 
I'll swap over to this slide. 
They're actually the same data, but the one on the left is zoomed in. 
You can see that it's only going from 1.25 seconds for period to 1.35 seconds for period. 
And the one on the right is zoomed out between one and two seconds, so you can see kind of the scale of the data. 
And the thing that we found was that for each of the different drop angles that we measured there was some variation, we saw very slight increase in amount of time per period for the angle. 
But when we look at it from the kind of larger scale shown on the right, we can see that it's actually a very insignificant amount of time that it changed. 
And it turns out to be well within the like point-zero-four (0.04) second margin of error.

\textbf{P2. Judy:} From this like we determined that it does fit Galileo's model because you don't see a change in the time of period when the angle increases. 
And we felt pretty confident about that because by dividing the pendulum swings by five it really reduces human error and the minor differences you see on the right are just the result of whatever human error is kind of left over.
\end{displayquote}
The group's narrative of their inquiry reflected both values of honesty as well as an interpretive bias that, consistent with previous literature, seems to arise from their confirmation framing. 
This manifests in their conclusion, that the data fits Galileo's model, in Peter's inaccurate comment that the variation in their data is contained within their margin of error, as well as in Judy's willingness to dismiss the trend as caused by "whatever human error is kind of left over."

After some process-oriented questions, another student in the class directly asks about the observed variation in their data:
\begin{displayquote}
    \textbf{P13. Jen:} Yeah, I just wanted to ask, cause like the times are super consistent but like when you do zoom in there is like a slight increase that seems to be like proportional with the increase in the angle, like do you guys have any ideas on maybe why that's happening or yeah?

    \textbf{P14. Peter:} To be honest, not really. 
    I think we talked about it kind of in depth while we were going, because like it seems too consistent to be ignored. 
    And we're assuming that there's something about the way that we're measuring it that was making the angle matter just a little bit. 
    Whether it was way we were dropping it or how exactly we were timing the end of the periods. 
    But we assumed that, since we're recording so many periods, right, that, that the difference between those is so small that it doesn't actually really matter, because it's within the margin of error that we've calculated.
\end{displayquote}
Like Judy in line 8, Jen highlights the seeming clarity of the data – it looks to be linear or proportional – as evidence for something going on. 
Peter's response mirrors the trajectory of their work as a group: First, Peter foregrounds honesty, admitting that they do not have an explanation and affirming that this is a good question. 
He even invokes a similar phrasing to Judy's original critique: "[the data] seems too consistent to be ignored" (P14). 
He continues by returning to human error as the potential cause:  they are assuming it's human error. 
Finally, he repeats their incorrect uncertainty analysis and states that the variation is just noise, and so an explanation is not necessary. 

Overall, this response makes explicit the deliberation and choices underlying their explanation. 
They show both graphs and identify the assumptions within their interpretation of the data. 
At the same time, in that interpretation there is evidence a confirmation framing led to questionable research practices. 

\subsection{Episode Summary}
Holly, Judy, and Peter view their data as a genuine problem and work to resolve it. 
In the beginning, all three participants signal, negotiate, and experience their efforts as productive inquiry. 
They engage in both epistemic and regulative actions as they develop ideas that help guide the direction of their intellectual work. 
In addition to the social negotiation occurring here, the apparatus, phenomenon, and data – in unprocessed and processed forms – are central to the emergence of epistemic agency. 

The material resistances they encounter motivate their intentional, disciplinary accommodations \cite{pickering_mangle_1995}. 
Because of this material context, their disciplinary practices serve a purpose and are understood by the students as useful for their inquiry. 
That they frame the activity as confirming a known result is a key reason that they seek out such accommodations. 
Holly and Peter's problematizing and troubleshooting work here involves grappling with and deconstructing the production of their data, a productive epistemological framing \cite{hardy_data_2020} that is nested within, and we suggest generated by, their confirmation framing of the lab as a whole.

In addition to the productive troubleshooting suggested by Smith \textit{et al.} \cite{smith_how_2020}, the students also create, refine, and make sense of concretized conceptual artifacts. 
This work includes attending to the procedural details of graph formatting as well as more conceptually substantive graphical analysis. 
Throughout, they wrestle with the entanglement of human and material agencies that facilitated the capture of this phenomenon. 

As the episode progresses, however, a shift occurs and the students' confirmation framing also manifests in questionable, biased interpretations of their data. 
It is notable that such a shift in activity occurs while they maintain confirmation framing. 
There are other aspects of their framing and group dynamics that do shift; Judy's fading participation as is meaningful shift in the group dynamics. 
Still, even as Judy participates less, Holly and Peter continue to demonstrate elements of intellectual honesty and epistemic accountability in their work – Peter even makes an explicit effort to involve Judy in their discussion and decision-making.

There are structural features that we expect to make a significant difference in their behavior, like Judy's remote participation and the wording of the instructions; at the same time, there are certain details, like the minor formatting mistake that Judy makes and how she and her groupmates react to it, that play as significant a role in this episode. 
The shift(s) that occurs in this episode highlights the complexity of social and epistemic dynamics in learning environments. 
The unfolding of this episode provides some generalizable insight into group dynamics and, at the same time, emerges from idiosyncrasies of this particular group in this particular context.

\section{Discussion}
\label{discussion}
Holly, Judy, and Peter's encounter with anomalous data – more specifically, the inconsistency between their results and their expectations – leads them to problematize, troubleshoot their apparatus, and produce various plots to analyze their data. 
In their problematizing, troubleshooting, and data interpretation, they are epistemically agentive: they take their data seriously as a meaningful reflection of the phenomenon they have constructed and, upon encountering unexpected results, enact various (epistemic) actions and disciplinary practices. 
Their confirmation framing supports this extended work to build an explanation for their discrepant data. 

That confirmation framing not only affords but supports productive disciplinary practices is in tension with previous findings that confirmation framing undermines epistemic agency \cite{stein_confirming_2018, smith_how_2020, phillips_not_2021}. Later in the episode, the students shift into activity more in line with those previous observations, with confirmation framing leading to questionable research practices and interpretive bias. 

The questionable research practices observed here resemble those reported in Smith \textit{et al.} \cite{smith_how_2020}. 
In both cases, students explicitly expect to confirm Galileo's model, that the periods should only depend on length, and yet encounter data that suggests otherwise – but not definitively – and assume that they are at fault. 
In Smith \textit{et al.} \cite{smith_how_2020}, students express clear discomfort with the middle ground, data that neither proves nor falsifies Galileo's model. 
In this case, Holly, Judy, and Peter do not consider it as an option, likely in part for the instructions emphasizing a binary choice. 
That contrasts with the instructions for students in the Smith \textit{et al.} \cite{smith_how_2020}, which explicitly recognize an indeterminate finding as valid. 

Notably, while many of the students in Smith \textit{et al.} \cite{smith_how_2020} express a "desire to be done," what Phillips \textit{et al.} \cite{phillips_not_2021} call a "hoops frame," we do not notice any apathy or even much off-topic chat among Holly, Judy, and Peter. 
Within their extended work of trying to figure out this problem and explain their data, they engage in multiple activities: troubleshooting, producing accurate graphs, data analysis, and graphical analysis. 

Moreover, Judy's comment in line 8 is unique across these accounts: her bid for the data reflecting a physical phenomenon does not invoke notions of error or wrongness. 
Judy emphasizes that the data has something to say. 
We argue that this move establishes or contributes to an expectation that their explanation of their data, whether for or against Galileo's model, is accountable to the data. 

That Judy is skeptical of Peter's initial conceptual explanation also implies that their explanation for this data needs to convince her as well. 
Not only do Peter and Holly seek to explain the specific trend that Judy focuses on (lines 20 and 24), but also Peter clearly seeks Judy's input and agreement for their final explanation. 
That said, Judy's responses indicate her exclusion from at least part of the conversation, in particular when she is unable to see the graph that Holly and Peter are discussing. 

More generally, throughout the episode, the tone and pace of conversation between Holly and Peter is different than when Judy participates.  
It is clear – and unsurprising – that Judy's not being in the room with Holly and Peter generates different experiences. 
In essence, we both see how intersubjectivity drives their enactment of (shared) epistemic agency and we see how constraints on that sharedness distort or inhibit epistemic agency.

\section{Implications}
\label{Implications}

We have identified several aspects of this learning environment that shaped their behavior:  supportive social interactions and access to them, values of intellectual honesty and epistemic accountability, material resistances that are clear to the students, the freedom to create knowledge-objects, as well as their framing the activity as confirming a known result. 
The last may be surprising, as confirmation framing is generally associated with limited epistemic agency \cite{phillips_not_2021, smith_how_2020}.

Here, as in other cases (e.g. \cite{brewe_modelling_2018, jeon_problematizing_2023, sundstrom_instructing_2023}) much seems to depend on the particular dynamics of the group and the context. 
For example, we suggest that Judy's remote participation encouraged her to focus on the data rather the apparatus, while just minutes later, it inhibited her participation. 
More striking, though, is what came of the minor formatting error she happened to make: prior to that moment, Judy had had a significant influence on the group's inquiry; after, she did not. 
We see it as an example of how idiosyncrasies can arise and be consequential in the complex dynamics of student inquiry. 
It suggests limits on what curriculum and course designs can accomplish in themselves to support students' epistemic agency.

Smith \textit{et al.} \cite{smith_how_2020} describe how the TA of the lab they analyzed sought to establish a norm for evidence-grounded claims and conclusions, similar to what we claim Judy does: 
\begin{displayquote}
"you all need to convince me that either there's a difference that you've measured between 10 and 20 degrees of the pendulum or there is no such difference that you can measure. I want you to sell me the package. Is there a difference between 10 and 20 degrees of the period of the pendulum? Alright? And you can do that by whatever means you see necessary" (p. 10).
\end{displayquote}
This move by the TA is qualitatively different than Judy's comment, but that is the point: his move positions him as the epistemic authority, the arbiter that the students have to convince. 
Judy occupies a different role in the classroom environment than the TA and her insistence on epistemic accountability is more subtle than this TA's attempt. 

Trends in education research and contemporary national curricular standards have shifted toward objectives of students' doing science, seeking to "engage students in knowledge construction – to position them as doers of science, rather than receivers of facts" (\cite{miller_addressing_2018} p. 1056). 
Yet, designing for doing science and effectively supporting students enacting epistemic agency is not simple \cite{sundstrom_problematizing_2020, phillips_not_2021, sundstrom_instructing_2023}
As more undergraduate science labs seek to promote disciplinary practices and epistemic agency, it is crucial to examine the dynamics underpinning the emergence of productive behavior.

\acknowledgments{We would like to thank Mark Akubo, Rebeckah Fussell, and Meagan Sundstrom for helping workshop the initial analysis of this episode. We also thank Julia Gouvea and Robert Hayes for thoughtful questions and generative conversations. This work is supported by the National Science Foundation, Grant DUE-2000394.}

\bibliography{references}
\end{document}